\documentclass[10pt,letterpaper]{article} 
\usepackage{opex3}
\usepackage{amsmath}
\usepackage{amssymb,graphicx}
\usepackage{cite}
\usepackage{bm}

\begin{document}

\newcommand{\vett}[1]{\mathbf{#1}}
\newcommand{\uvett}[1]{\hat{\vett{#1}}}
\newcommand{\beq}{\begin{equation}}
\newcommand{\eeq}{\end{equation}}
\newcommand{\barr}{\begin{eqnarray}}
\newcommand{\earr}{\end{eqnarray}}


\title{Radially and azimuthally polarized nonparaxial Bessel beams made simple}

\author{Marco Ornigotti$^{1,*}$ and Andrea Aiello$^{1,2}$}
\address{$^{1}$Max Planck Institute for the Science of Light, G$\ddot{u}$nther-Scharowsky-Strasse 1/Bau24, 91058 Erlangen, Germany\\
$^{2}$Institute for Optics, Information and Photonics, University of Erlangen-Nuernberg, Staudtstrasse 7/B2, 91058 Erlangen, Germany}
\email{*marco.ornigotti@mpl.mpg.de}

\begin{abstract}
	We present a method for the realization of radially and azimuthally polarized nonparaxial Bessel beams in a rigorous but simple manner. This result is achieved by using the concept of Hertz vector potential to generate exact vector solutions of Maxwell's equations from scalar Bessel beams. The scalar part of the Hertz potential is built by analogy with the paraxial case as a linear combination of Bessel beams carrying a unit of orbital angular momentum. In this way we are able to obtain spatial and polarization patterns analogous to the ones exhibited by the standard cylindrically polarized paraxial beams. Applications of these beams are discussed.
\end{abstract}



\section{Introduction}
 Among the broad zoology of solutions of Maxwell's equations, radially and azimuthally polarized vector beams, i.e., beams of light with non uniform polarization patterns, attracted great interest in the last decade\cite{ref1,ref2,ref3}. Their particular character, in fact, makes them very appealing for applications in various fields of research such as single molecule spectroscopy \cite{ref4}, optical tweezing \cite{ref5}, confocal microscopy \cite{ref6} as well as in commercial applications like material processing \cite{ref7,ref7a}. Another interesting feature of such vector beams is that they can yield  to very small focal spots \cite{ref10} and generate axial electric \cite{ref11} or magnetic \cite{ref12} fields once focused. Radially and azimuthally polarized paraxial fields have also been investigated in uniaxial crystals both theoretically \cite{ale1,referee12} and experimentally \cite{referee11}. Moreover, the interest for these complex polarization field configurations does not limit only to the linear optics case, as azimuthally polarized spatial dark solitons have been proved to be exact solutions of Maxwell's equations in Kerr-type nonlinear media \cite{ale2}.  All these reasons have motivated the development of several experimental techniques to realize such field configurations  \cite{ref13,ref14,ref15,ref16,ref16a}. Recently, it has been also shown that squeezed azimuthally polarized optical beams show hybrid entanglement \cite{ref17}.  For a comprehensive theoretical analysis of the properties (both classical and quantum) of general cylindrically polarized states of light, the reader is addressed to \cite{ref18}. 
 
 Despite the considerable amount of work that has been done in this field, radially and azimuthally polarized beams have been studied almost exclusively within the framework of paraxial optics. Although the study of strongly focused paraxial beams gave some insight on the nonparaxial properties of light \cite{ref18a}, and some attempt to a rigorous description of non paraxial beams has been given using complex dipole sources and sinks \cite{shepard1}, perturbative corrections of Gaussian beams \cite{salmon} and non paraxial elegant Laguerre-Gauss beams \cite{april1}, the sole theoretical work on nonparaxial Bessel beams, to the knowledge of the authors, is represented by the work of Bouchal and Olivik \cite{vectorBeams}, where the concept of non-diffractive vector beam is introduced and extended to the nonparaxial domain. However, we feel that the physical meaning of the solutions found in  \cite{vectorBeams} tends to be obscured by the use of a somehow heavy mathematical formalism.  It is our aim here to introduce cylindrically polarized states of light beyond the paraxial domain in a much simple way, by analogy with the paraxial case which leads to a clearer physical understanding, thus making this important subject accessible to a broader audience. Thanks to the fact that we use Bessel beams as basis to create these new fields, they will also possess the typical characteristics of Bessel beams, such as self healing and diffractionless propagation. They could also be used, as it has been recently proposed \cite{OL2012}, to generate accelerating beams of arbitrary trajectories. All these features, combined with the complex cylindrical polarization patterns introduced here, may lead to new possible developments in timely research fields such as strongly focused light or to investigate the role of complex polarization patterns in light-atom scattering problems \cite{nostroFuturo}.
 
\section{Radially and azimuthally polarized vector Bessel beams}

As shown in  \cite{ref18}, in the paraxial case radially and azimuthally polarized beams live in a four dimensional space spanned by the basis formed by the Cartesian product of the base of Hermite-Gauss modes $\psi_{nm}(\vett{r})$ of order $N=n+m=1$ \cite{padgett} $\{\psi_{10},\psi_{01}\}$ and the polarization basis $\{\uvett{x},\uvett{y}\}$, namely $\{\psi_{10},\psi_{01}\}\otimes\{\uvett{x},\uvett{y}\}=\{\psi_{10}\uvett{x},\psi_{10}\uvett{y},\psi_{01}\uvett{x},\psi_{01}\uvett{y}\}$. Linear combinations of these four vectors give rise to radially ($\uvett{u}_R$) and azimuthally ($\uvett{u}_A$) polarized fields as follows:
\begin{subequations}\label{basi}
\begin{align}
\uvett{u}_R^{\pm}=\frac{1}{\sqrt{2}}(\pm\psi_{10}\uvett{x}+\psi_{01}\uvett{y}),\\
\uvett{u}_A^{\pm}=\frac{1}{\sqrt{2}}(\mp\psi_{01}\uvett{x}+\psi_{10}\uvett{y}),
\end{align}
\end{subequations}
where the $\pm$ sign indicates co-rotating and counter-rotating modes respectively \cite{ref18}.

It is moreover well known, in paraxial optics, that suitable linear combinations of Hermite-Gaussian modes may lead to orbital angular momentum carrying beams \cite{woerdman}. In particular, it is possible to demonstrate that the Hermite-Gauss modes $\psi_{10}$ and $\psi_{01}$ can be written as a superposition of two Laguerre-Gaussian beams with angular momentum $m=\pm 1$ \cite{padgett}. Having this in mind,  we now construct the nonparaxial counterpart of the Hermite-Gaussian basis modes, using combinations of vector Bessel beams in such a way that the angular momentum of the resulting nonparaxial beam is preserved.

 First, we choose for the scalar field a $m$th-order Bessel beam, namely 
 \beq
 \psi_m^B(\vett{r})=J_m(K_0 R)\exp{(im\phi)}\exp{(iz\sqrt{k^2-K_0^2})},
 \eeq
 where $K_0=k\sin\theta_0$ (being $\theta_0$ the cone angle aperture of the Bessel beam), $x=R\cos\phi$ and $y=R\sin\phi$. Henceforth with $J_m(x)$ we denote the $m$ order bessel function.
 
 Then we construct, by analogy with the paraxial case, the nonparaxial scalar beams $\Psi_{10}(\vett{r})$ and $\Psi_{01}(\vett{r})$ by combining two Bessel beams $\psi_m^B(\vett{r})$ with angular momentum $m=\pm 1$ respectively, obtaining 
\begin{subequations}\label{connection}
\beq
 \Psi_{10}(\vett{r})=\frac{1}{\sqrt{2}}[\psi_1^B(\vett{r})+\psi_{-1}^B(\vett{r})],
\eeq
\beq
\Psi_{01}(\vett{r})=-\frac{i}{\sqrt{2}}[\psi_1^B(\vett{r})-\psi_{-1}^B(\vett{r})].
\eeq
\end{subequations}
Before going any further, it is worth discussing the paraxial limit of these two scalar fields $\Psi_{10}(\vett{r})$ and $\Psi_{01}(\vett{r})$, in order to prove that they reduce correctly to the Hermite-Gauss beams $\psi_{10}(\vett{r})$ and $\psi_{01}(\vett{r})$. In the paraxial limit $K_0=k\sin\theta_0\simeq k\theta_0=2/w_0$, where $w_0$ is the beam waist, and $k_z\simeq k$. Substituting these expressions into Eqs. \eqref{connection} and expanding in a Taylor series for $w_0\rightarrow\infty$ (the paraxial limit corresponds to small propagation angles with respect to the main propagation direction or, equivalently, to large beam waists since $w_0\propto1/\theta_0$) gives
\begin{subequations}
\barr
\Psi_{10}(\vett{r})\simeq\frac{\sqrt{2\,}x}{w_0}e^{ikz}\left(1-\frac{x^2+y^2}{2w_0^2}+ \mathcal O(\theta_0^4)\right)\simeq\frac{\sqrt{2}\,x}{w_0}e^{ikz}e^{-\frac{x^2+y^2}{2w_0^2}},\\
\Psi_{01}(\vett{r})\simeq\frac{\sqrt{2}\,y}{w_0}e^{ikz}\left(1-\frac{x^2+y^2}{2w_0^2}+ \mathcal O(\theta_0^4)\right)\simeq\frac{\sqrt{2}\,y}{w_0}e^{ikz}e^{-\frac{x^2+y^2}{2w_0^2}}.
\earr
\end{subequations}
As can be seen, the paraxial limit of the quantities $\Psi_{10}(\vett{r})$ and $\Psi_{01}(\vett{r})$ gives exactly the Hermite-Gauss beams $\psi_{10}(\vett{r})$ and $\psi_{01}(\vett{r})$ respectively since, $\psi_{10}(\vett{r})\propto x \exp{[-(x^2+y^2)/2w_0^2]}$. and $\psi_{01}(\vett{r})\propto y\exp{[-(x^2+y^2)/2w_0^2]}$ \cite{svelto}. 

\begin{figure}[!t]
\begin{center}
\includegraphics[width=0.8\textwidth]{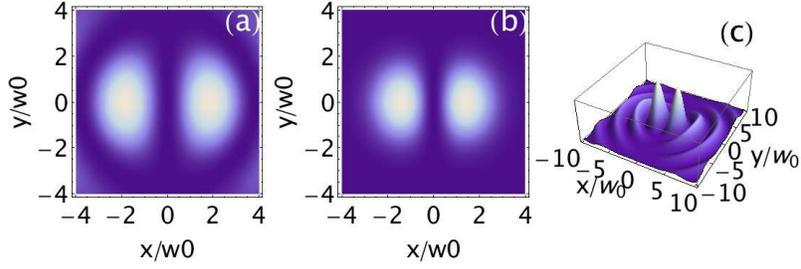}
\caption{ (a) Contour plot of the scalar function $\Psi_{10}(\vett{r})$ close to the propagation axis. (b) Contour plot of the Hermite-Gauss beam $\psi_{10}$. (c) Three dimensional intensity profile of the scalar function $\Psi_{10}$. A direct comparison between panels (a) and (b) shows that in the paraxial domain $\Psi_{10}$ correctly reproduces the behavior of the Hermite-Gauss beam $\psi_{10}$. Note moreover that from panel (c)  for $x/w_0\geq 4$ and $y/w_0\geq 4$, the scalar function $\Psi_{10}(\vett{r})$ shows some ripples, whose intensity is much smaller than the central lobes.}
\label{figura_zero}
\end{center}
\end{figure}

In Fig. \ref{figura_zero} the intensity profile of the scalar beam of $\Psi_{10}(\vett{r})$ is plotted as a function of the dimensionless coordinates $x/w_0$ and $y/w_0$ being $w_0=2/(k\theta_0)$ the beam waist. From this figure it is clear that, if we limit ourselves to the central domain [Fig. \ref{figura_zero} (a)], i.e. if we consider only the paraxial limit,  the intensity distribution of the nonparaxial beam $\Psi_{10}(\vett{r})$ is fully equivalent to the intensity distribution of the correspondent Hermite-Gauss beam $\psi_{10}$.

Finally, following  Eqs. \eqref{basi} we build the nonparaxial radially and azimuthally basis as
 \begin{subequations}\label{nonparaxial_basis}
 \barr
 \uvett{U}_R^{\pm}&=&\frac{1}{\sqrt{2}}\left(\pm\Psi_{10}\uvett{x}+\Psi_{01}\uvett{y}\right ),\\
 \uvett{U}_A^{\pm}&=&\frac{1}{\sqrt{2}}\left (\mp\Psi_{01}\uvett{x}+\Psi_{10}\uvett{y}\right ).
 \earr
 \end{subequations}

These modes are \emph{exact} vector solutions of the full Helmholtz equation, since the scalar fields $\Psi_{10}(\vett{r})$ and $\Psi_{01}(\vett{r})$ are linear combinations of solutions of the full Helmholtz equation themselves, in contrast with the case of Hermite-Gaussian beams that are solution of the paraxial equation solely.  However, they are not exact solutions of Maxwell's equations, as they do not evidently satisfy the transversality condition $\nabla\cdot\uvett{U}_{R,A}^{\pm}=0$. 

However, because of these features, we can regard the modes $\vett{U}^{\pm}_{R,A}$ as bona fide Hertz vector potentials for the electromagnetic field \cite{jackson}. Exact solutions of the Maxwell's equations are obtained by choosing either $\boldsymbol{\Pi}(\vett{r},t)=\vett{U}^{\pm}_{R}\exp(-i\omega t)$ or $\boldsymbol{\Pi}(\vett{r},t)=\vett{U}^{\pm}_{A}\exp(-i\omega t)$ in the following equations:

\begin{subequations}
\barr
\vett{E}(\vett{r},t)&=&\bf{\nabla}\times\left(\bf{\nabla}\times\boldsymbol{\Pi}\right),\label{hertz}\\
\vett{B}(\vett{r},t)&=&\frac{1}{c^2}\frac{\partial}{\partial t}\left(\bf{\nabla}\times\boldsymbol{\Pi}\right).\label{hertz_mag}
\earr
\end{subequations}

In the general case, two independent potentials are introduced: the Hertz electric and magnetic vector potentials $\boldsymbol{\Pi}_e$ and $\boldsymbol{\Pi}_m$, respectively. The difference between $\boldsymbol{\Pi}_e$ and $\boldsymbol{\Pi}_m$ resides in the sources of external electric and magnetic polarization densities $\mathbf{P}_\text{ext}$ and $\mathbf{M}_\text{ext}$, respectively, that generate them \cite{jackson}.  Anyhow,  since we are considering electromagnetic fields in vacuum, $\mathbf{P}_\text{ext}=0=\mathbf{M}_\text{ext}$ and a single Hertz vector potential $\boldsymbol{\Pi}$ can be used. According to the convention adopted by Jackson, our $\boldsymbol{\Pi}$ coincides with $\boldsymbol{\Pi}_e$ in vacuum. From Eq. \eqref{hertz} we obtain the full electric vector field for the radially and azimuthally polarized nonparaxial beams, whose components, in cylindrical coordinates $\{R,\theta,z\}$, read explicitly as follows:
\begin{subequations}\label{radp}
\barr
E^r_{R+}(\vett{r})&=&k_z^2J_1(K_0R)e^{ik_z z},\\
E^{\theta}_{R+}(\vett{r})&=& 0,\\
E^z_{R+}(\vett{r})&=&i k_z K_0J_0(K_0R)e^{ik_z z},
\earr
\end{subequations}
for the co-rotating radially polarized field,
\begin{subequations}\label{radm}
\barr
E^r_{R-}(\vett{r})&=&\frac{\cos(2\theta)}{R^2}\Big[2K_0RJ_0(K_0R)-J_1(K_0R)(4+k_z^2R^2)\Big]e^{ik_z z},\\
E^{\theta}_{R-}(\vett{r})&=&\frac{\sin(2\theta)}{R^2}\Big[2K_0RJ_0(K_0R)-J_1(K_0R)(4-k^2R^2)\Big]e^{ik_z z},\\
E^z_{R-}(\vett{r})&=&i k_z K_0\cos (2\theta) J_0(K_0R)e^{ik_z z},
\earr
\end{subequations}
for the counter-rotating radially polarized field,
\begin{subequations}\label{azp}
\barr
E^r_{A+}(\vett{r})&=&0,\\
E^{\theta}_{A+}(\vett{r})&=&k^2J_1(K_0R)e^{ik_z z},\\
E^z_{A+}(\vett{r})&=&0,
\earr
\end{subequations}
for the co-rotating azimuthally polarized field, and
\begin{subequations}\label{azm}
\barr
E^r_{A-}(\vett{r})&=&\frac{\sin(2\theta)}{R^2}\Big[-2K_0RJ_0(K_0R)+J_1(K_0R)(4+k_z^2R^2)\Big]e^{ik_z z},\\
E^{\theta}_{A-}(\vett{r})&=&\frac{\cos(2\theta)}{R^2}\Big[2K_0RJ_0(K_0R)-J_1(K_0R)(4-k^2R^2)\Big]e^{ik_z z},\\
E^z_{A-}(\vett{r})&=&-ik_zK_0\sin(2\theta) J_2(K_0R)e^{ik_z z},	
\earr
\end{subequations}
for the counter-rotating azimuthally polarized field. Here, $\vett{r}=\{R\, \uvett{R},\theta\, \hat{\boldsymbol{\theta}},z\, \uvett{z}\}$, and $k_z=\sqrt{k^2-K_0^2}$. The expression of the co-rotating fields are in accordance with the one presented in  \cite{vectorBeams}, section 2 examples 1 and 2. From Eq. \eqref{hertz_mag}, in a similar manner we can obtain the expression for the magnetic field:
\begin{subequations}
\barr
cB^r_{R+}(\vett{r})&=&0,\\
cB^{\theta}_{R+}(\vett{r})&=&kk_zJ_1(K_0R)e^{ik_z z},\\
cB^z_{R+}(\vett{r})&=&0,
\earr
\end{subequations}
for the co-rotating radially polarized field,
\begin{subequations}
\barr
cB^r_{R-}(\vett{r})&=&-kk_z\sin (2\theta) J_1(K_0R)e^{ik_z z},\\
cB^{\theta}_{R-}(\vett{r})&=&-kk_z\cos (2\theta) J_1(K_0R)e^{ik_z z},\\
cB^z_{R-}(\vett{r})&=&ikK_0\sin (2\theta) J_2(K_0R)e^{ik_z z},
\earr
\end{subequations}
for the counter-rotating radially polarized field,
\begin{subequations}
\barr
cB^r_{A+}(\vett{r})&=&-kk_zJ_1(K_0R)e^{ik_z z},\\
cB^{\theta}_{A+}(\vett{r})&=&0,\\
cB^z_{A+}(\vett{r})&=&-ikK_0J_0(K_0R)e^{ik_z z},
\earr
\end{subequations}
for the co-rotating azimuthally polarized field, and
\begin{subequations}
\barr
cB^r_{A-}(\vett{r})&=&-kk_z\cos (2\theta) J_1(K_0R)e^{ik_z z},\\
cB^{\theta}_{A-}(\vett{r})&=&-kk_z\sin (2\theta) J_1(K_0R)e^{ik_z z},\\
cB^z_{A-}(\vett{r})&=&ikK_0\cos (2\theta) J_2(K_0R)e^{ik_z z},
\earr
\end{subequations}
for the counter-rotating azimuthally polarized field.
\begin{figure}[!t]
\begin{center}
\includegraphics[width=0.8\textwidth]{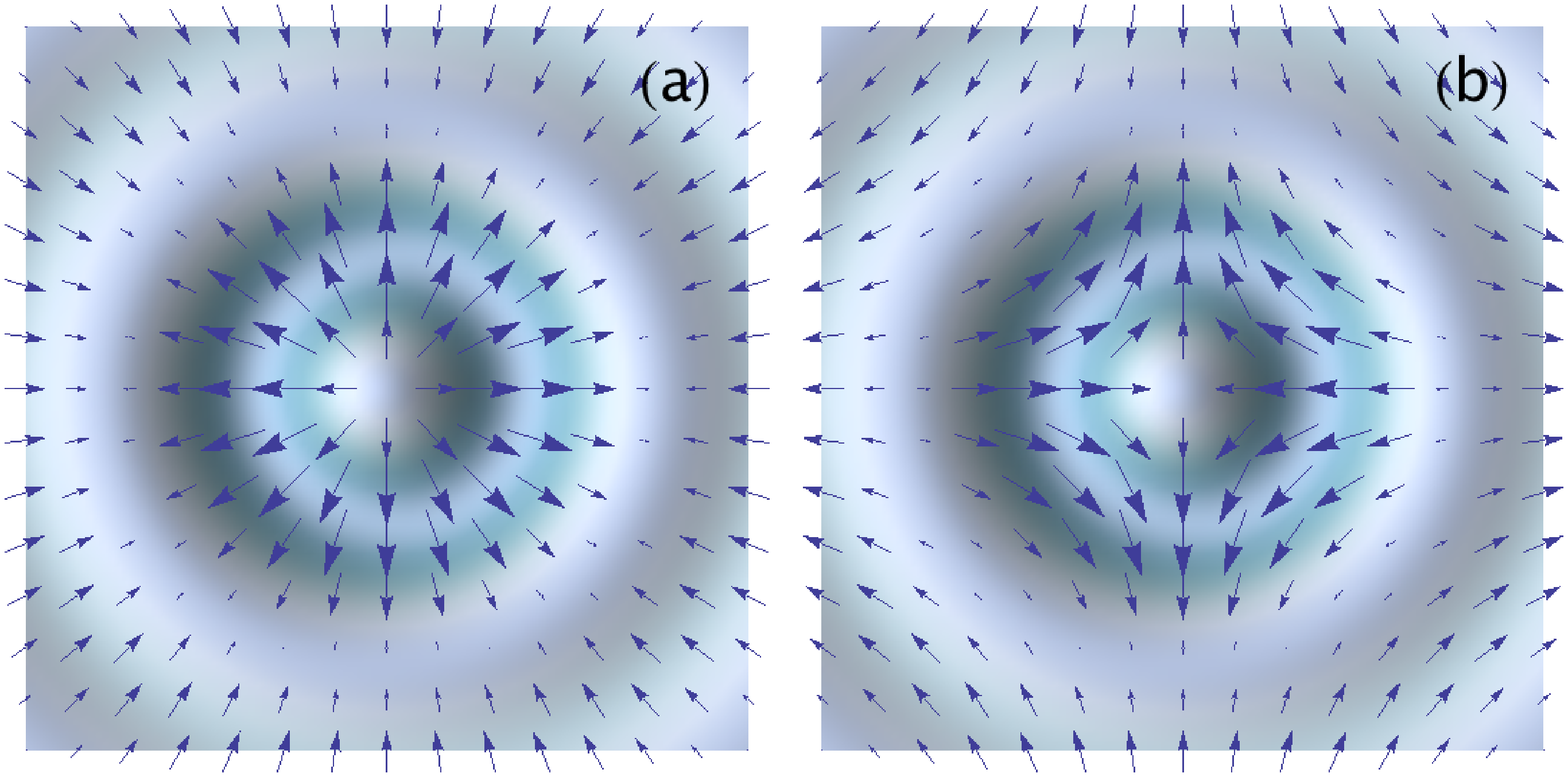}
\caption{Complex polarization patterns of (a) co-rotating radially polarized electric field $\vett{E}^+_R$ and (b) counter-rotating radially polarized electric field $\vett{E}^-_R$, with superimposed the donut-shaped intensity distribution. The axes of both graphs span the interval $[-5,5]$ in units of the beams waist $w_0$.}
\label{figure1}
\end{center}
\end{figure}
\begin{figure}[!t]
\begin{center}
\includegraphics[width=0.8\textwidth]{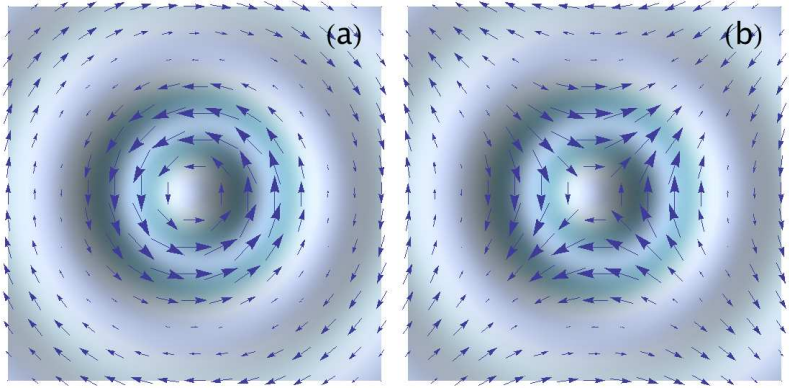}
\caption{Complex polarization patterns of (a) co-rotating azimuthally polarized electric field $\vett{E}^+_A$ and (b) counter-rotating azimuthally polarized electric field $\vett{E}^-_A$,with superimposed the donut-shaped intensity distribution. The axes of both graphs span the interval $[-5,5]$ in units of the beams waist $w_0$.}
\label{figure2}
\end{center}
\end{figure}

The polarization pattern as well as the intensity distributions for the electric field of these nonparaxial radially and azimuthally polarized beams are displayed in Figs. \ref{figure1} and \ref{figure2}.  

\begin{figure}[!t]
\begin{center}
\includegraphics[width=0.5\textwidth]{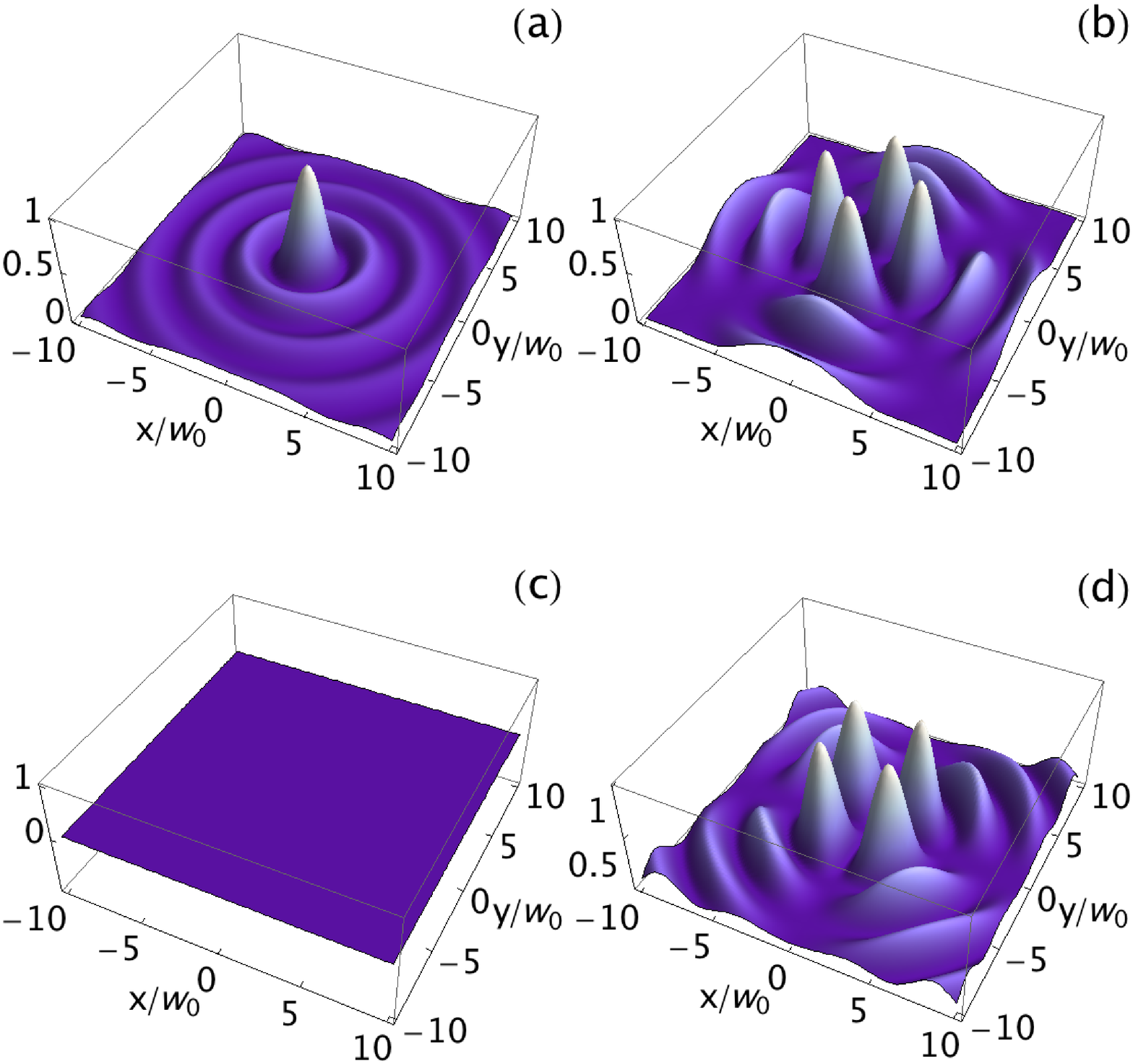}
\caption{Normalized intensity distributions for the longitudinal component $E_z$ of the radially (upper row) and azimuthally (lower row) polarized nonparaxial fields. Panels (a)-(c) and (b)-(d) display the co-rotating and counter-rotating fields respectively.}
\label{figure3}
\end{center}
\end{figure}

 As can be seen from Eqs. \eqref{radp}-\eqref{azm}, with exception of the co-rotating azimuthally polarized beams [Eqs. \eqref{azp}], these fields present a nonzero longitudinal component (due to their nonparaxial nature), whose intensity distribution is shown in Fig. \ref{figure3} for the case of radial polarization (upper panel) and azimuthal polarization (lower panel).

\section{Discussion}
Bessel beams are a versatile and powerful tool for theoreticians. They are \emph{exact} solution of the scalar Helmholtz equation and they allow a fully analytic treatment of the electromagnetic problems in which they are involved. Unfortunately, just like the more ubiquitous plane waves, Bessel beams carry infinite energy, a feature that makes them be unphysical. However, when calculating physical quantities possessed by Bessel beams such as angular momentum \cite{angularBessel} or beam shifts \cite{shiftBessel} (to name a few), this difficulty can be easily removed by using a proper normalization \cite{haus}. Another way of removing this apparent unphysical character of Bessel beams is to use their regularized (and then experimentally realizable) versions: the so-called Bessel-Gauss beams \cite{nuovo1,nuovo2}, which possess a finite energy but, on the other side, they cannot be represented with a simple closed form analytic expression, as they can be written only as infinite series. Moreover, while Bessel beams are diffractionless (due to their singular angular spectrum), Bessel-Gauss beams are not, as they maintain the diffraction less character only up to a certain critical length. Despite these differences, the formalism presented here for Bessel beams can be straightforwardly extended to the case of Bessel-Gauss beams, allowing a more experimentally feasible framework to generate non paraxial cylindrically polarized light fields.

\section{Conclusions}

In conclusion, we have shown how to generalize the theoretical framework of radially and azimuthally polarized beams to the case of nonparaxial fields, using Bessel beams as generating Hertz vector potentials. We shown that these nonparaxial fields retain the typical donut-shape of their paraxial counterparts, although they show a nonzero longitudinal electric field component. 

In conclusion, we stress the fact that according with the definition given in  \cite{ref18}, radially and azimuthally polarized beams built with the rule given by Eq. \eqref{nonparaxial_basis} (or its paraxial counterpart given by Eq. \eqref{basi}) by combining beams with one unit of angular momentum are the only one that generate beams with \emph{total} (spin plus orbital) angular momentum equal to zero. Therefore, an extension of this formalism to higher values of orbital angular momentum by keeping constant the total angular momentum, is not possible. However if one relaxes this assumption, cylindrically polarized with higher orbital angular momentum, can be generated by using $q$-plates \cite{lorenzo,lorenzo2}.

\end{document}